# Gap soliton gating, dissociation, and retrieval via defect mode excitation in a resonant photonic crystal

Igor V. Mel'nikov[1] and J. Stewart Aitchison

*The E. S. Rogers Sr. Department of Electrical and Computer Engineering, University of Toronto,
10 King's College Road, Toronto, Ontario, Canada M5S 3G4*

**Abstract:** *We consider the interaction within a bunch of two coherent optical pulses in a one dimensional, resonant photonic crystal which includes a defect produced by a coherent pump. Using numerical methods, the defect is shown to provide a selective reversible trap for a gap soliton mediated by the long-range coupling. This opens up new opportunities for signal transmission control and light localization.*

There has been a recent resurgence of interest in photonic crystals (PC) with resonant nonlinearity, mainly driven by dramatic advances in manufacturing InGaAs/GaAs Bragg periodic multiple quantum well structures and rare-earth doped AlGaAs/GaAs nanostructures [1-4]. This suggest that such building block of future optical computing as gap soliton (GS) might be experimentally feasible at moderate level of light intensity ~ 10 MW/cm$^2$. Recently, we have proposed an effective mean of control over these solitons when a defect associated with a weak linear excitation or incoherent pump inside a resonant photonic crystal (RPC) modifies GS propagation [5-7]. It turns out that the periodicity breakup is unnecessary for the GS to be trapped inside the RPC because a weak non-propagating field creates a potential that acts as a phase-sensitive trap. In this Letter, using the results of a detailed and extensive theoretical study of excitation trapping inside RPCs [6,7], a further interesting consequences of inverting a small length inside a RPC is explored: all-optical filtering and memory cell effect.

First of all, relevant expressions from the theory are abstracted and adapted. Assuming a collinear, distributed feedback two-wave interaction in a medium of periodically positioned two-level atoms, the governing Maxwell-Bloch equations take the form [5-7]:

$$\Omega_t^\pm \pm \Omega_x^\pm = P, \tag{1a}$$

$$P_t = n(\Omega^+ + \Omega^-), \tag{1b}$$

$$n_t = -\mathrm{Re}\{P^*(\Omega^+ + \Omega^-)\}, \tag{1c}$$

where $\Omega^\pm$ is the normalized electric field of the pulse, $P$ is the resonant polarization, and $n$ is the population inversion. As input values, we use the following conditions:

---

1) corresponding author; now with Comtex Consulting Inc., 162 Princess Margeret Blvd., Etobicoke, Ontario, Canada M9B 2Z5; can be reached at phone (416) 946-5524, fax (416) 971-3020, or e-mail *igor.melnikov@utoronto.ca*

$$\Omega^+(x=0,t) = \Omega_1 \text{sech}[(t-t_1)/\tau_p] \pm \Omega_2 \text{sech}[(t-t_2)/\tau_p],$$

$$\Omega^-(x=L,t)=0, \quad \Omega^\pm(x,t=0)=0, \quad P(x,t=0)=0, \tag{2}$$

here $\tau_p$ is the normalized duration of the incident pulse.

A very useful quantity governs the nonlinear interaction between the GS and defect mode provided by the inversion inside the RPC:

$$\Phi(t) = \frac{\Gamma_0}{4} \int_{-\infty}^{+\infty} dx \, \text{sech}(\sqrt{2}x) \tilde{\Omega}(x,t), \tag{3}$$

where

$$\tilde{\Omega}(x,t) = \pm \frac{2\sqrt{2}}{\sqrt{1-u^2}} \text{sech}\left[\frac{\sqrt{2}(x-\xi(t))}{\sqrt{1-u^2}}\right]. \tag{4}$$

Here $u(t)=\xi_t$ is the velocity of the soliton, and the plus sign corresponds to the anti-kink whereas the minus gives the kink solution. It is seen that by means of the relevant choice of the parameters the potential (3) can be made large enough to exceed the kinetic energy, $u^2/2$, of the slow GS. More specifically, the defect mode inside the RPC reflects the GS with $\tilde{\Omega} > 0$ (anti-kink) and let that one with $\tilde{\Omega} < 0$ (kink) to pass.

This analysis is now used to explore the behavior of a pair of GSs in the presence of a defect mode that is due to the total population inversion in the middle of the sample. In Fig. 1, the projection of the population inversion is plotted onto an (*x, t*)-plane for an in-phase input (two kinks). The defect is transparent only within certain window of the time delay between the two GS; the longer delays yield the leading GS to be trapped at the defect and to close the transparency.

In Fig. 2 we contrast these results with the case of the interaction of the out-of-phase GSs. Clearly the attraction between the closely separated solitons surpasses the interaction of each of them with the defect, and formation of the breather-like bound state and its free traversing the defect are observed [Fig. 2 (a)]. As the delay time between the soliton and anti-soliton grows, the pattern discussed in our earlier publications [6,7] (and not shown here) emerges: the unstable breather is formed that is dissociated later on owing to the barrier created by the gain strip for the second pulse (anti-soliton). Further increase of the delay leads to a completely new and unexpected result [Fig. 2 (b,c)]: the stable breather is formed at and instantly escapes from the defect. This effect however, exists only for certain values of the delay, that is we observe an intermittent sequence of the breather dissociation and trapping when the second pulse, anti-soliton, is launched into the RPC at the times different than that of Fig. 2 (b,c). Notice that in the case of the breather buildup at and escape from the defect there is no energy left on or taken out from the inversion strip.

It is natural to assume that the dynamics of trapping of the GS may also be controlled by means of its interaction with a more intense GS. As a relevant example,

Fig. 3 shows the results of the numerical integration of the Maxwell-Bloch equations (1) for the case of the two solitons interaction with different input amplitudes. Above a certain threshold, the trailing GS with larger energy permits the escape of the GS from the trapping area; again, it may happen only within a certain delay window.

In conclusion, we demonstrate that using both analytical and numerical methods interesting and previously unforeseen properties of GSs in RPC can be predicted and explained in a physically transparent form. The most important result is the fact that we are able to show that the oscillating GS created by the presence of an inversion inside the RPC can be manipulated by means of a proper choice of bit rate, phase and amplitude modulation. Using this approach, we were able to obtain qualitatively different regimes of the RPC operation. A noticeable observation is that both the delay time and amplitude difference must exceed a certain level to ensure effective control over a soliton dynamics. The modification of the defect that accomplishes the soliton trapping can make the dynamics of soliton trains in the resonant photonic crystal with defect even more interesting and is a subject of the future work.

This study was supported by Natural Sciences and Engineering Research Council of Canada and Photonics Research Ontario..

# FIGURE CAPTIONS

**Fig. 1**  Repulsive interaction of the two in-phase gap solitons in thee presence of the inversion in the middle of the sample for three values the delay time

**Fig. 2**  Bound-state creation/dissociation for a pair of the out-of-phase gap soliton in the RPC in the presence of the inversion in the middle of the sample

**Fig. 3**  Memory cell – the GS is released after collision with more intense pulse.

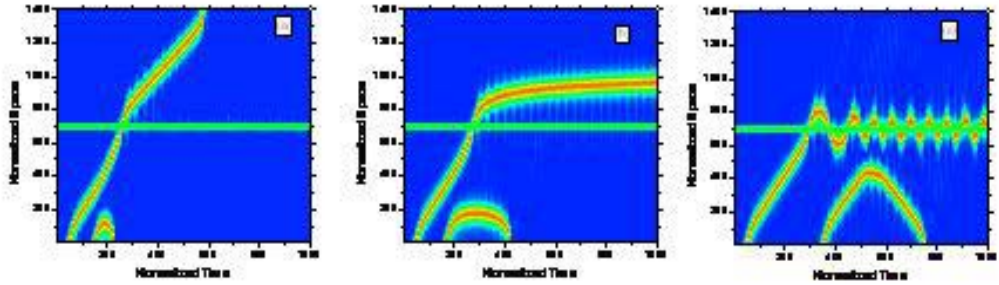

Mel'nikov and Aitchison, Fig. 1 of 3

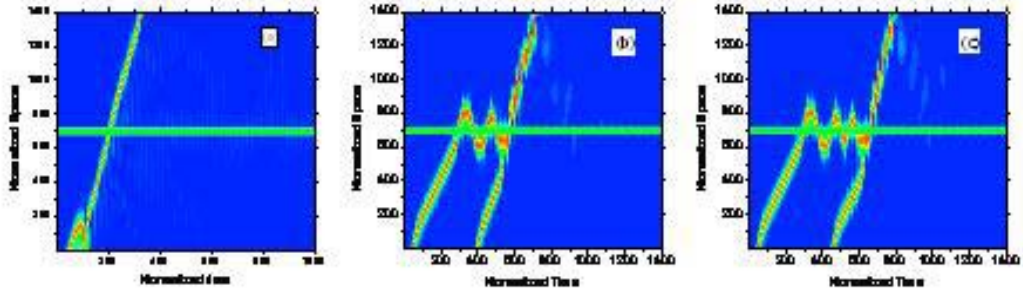

Mel'nikov and Aitchison, Fig. 2 of 3

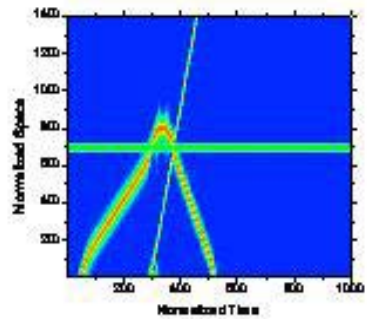

Mel'nikov and Aitchison, Fig. 3 of 3

5